\documentclass[journal,10pt]{IEEEtran}
\pdfoutput=1
\makeatletter
\def\input@path{{./}{Paper/}}
\makeatother

\usepackage{amsmath,amsfonts,bm}

\def\1{\bm{1}}

\DeclareMathAlphabet{\mathsfit}{\encodingdefault}{\sfdefault}{m}{sl}
\SetMathAlphabet{\mathsfit}{bold}{\encodingdefault}{\sfdefault}{bx}{n}

\usepackage{times}
\usepackage[hypertexnames=false]{hyperref}
\usepackage{url}
\usepackage{booktabs}
\usepackage{amsmath,amssymb,amsfonts}
\usepackage{graphicx}
\usepackage{multirow}
\usepackage{algorithm}
\usepackage{algorithmic}
\usepackage{color}
\usepackage{tikz}
\usetikzlibrary{shapes,arrows,positioning,fit,calc,shadows,shapes.geometric,backgrounds,decorations.pathmorphing,arrows.meta}

\usepackage[numbers,sort&compress]{natbib}

\graphicspath{{./figures/}{Paper/figures/}}

\title{TimberAgent: Gram-Guided Retrieval for Executable \\  Music Effect Control}
\author{Shihao He*,  Yihan Xia*, Fang Liu, Taotao Wang, and Shengli Zhang\\
\thanks{S. He, Y. Xia, F. Liu,  T.~Wang and S. Zhang are with the College of Electronics and Information Engineering, Shenzhen University, Shenzhen, China, e-mails: 2410044045@mails.szu.edu.cn;\{xiayihan2023, liuf, ttwang, zsl\}@szu.edu.cn } 
\thanks{S. He and Y. Xia contributed equally to this work.  (Corresponding author: F. Liu and T. Wang)}}

\begin{document}
\newcommand{\paperbibliostyle}{IEEEtranN}
\maketitle

\begin{abstract}
Digital audio workstations expose rich effect chains, yet a semantic gap remains between perceptual user intent and low-level signal-processing parameters. We study retrieval-grounded audio effect control, where the output is an editable plugin configuration rather than a finalized waveform. Our focus is Texture Resonance Retrieval (TRR), an audio representation built from Gram matrices of projected mid-level Wav2Vec2 activations. This design preserves texture-relevant co-activation structure. We evaluate TRR on a guitar-effects benchmark with 1,063 candidate presets and 204 queries. The evaluation follows Protocol-A, a cross-validation scheme that prevents train-test leakage. We compare TRR against CLAP and internal retrieval baselines (Wav2Vec-RAG, Text-RAG, FeatureNN-RAG), using min-max normalized metrics grounded in physical DSP parameter ranges. Ablation studies validate TRR's core design choices: projection dimensionality, layer selection, and projection type. A near-duplicate sensitivity analysis confirms that results are robust to trivial knowledge-base matches. TRR achieves the lowest normalized parameter error among evaluated methods. A multiple-stimulus listening study with 26 participants provides complementary perceptual evidence. We interpret these results as benchmark evidence that texture-aware retrieval is useful for editable audio effect control, while broader personalization and real-audio robustness claims remain outside the verified evidence presented here.
\end{abstract}


\section{Introduction}
\label{sec:introduction}

	Neural audio synthesis faces a fundamental tension between \textit{fidelity} and \textit{control}. Recent high-fidelity generative models, primarily based on latent diffusion \citep{liu2023audioldm,agostinelli2023musiclm}, can achieve strong perceptual quality but often operate as ``black boxes.'' They generate finalized waveforms where internal attributes—such as the attack time of a compressor or the decay of a reverb—are entangled and inaccessible for post-hoc editing. Conversely, Differentiable Digital Signal Processing (DDSP) \citep{engel2020ddsp} offers interpretable, structured control but often suffers from the ill-posed nature of the inverse parameter estimation problem.

This tension motivates the development of retrieval-based audio control systems that can map perceptual intent to interpretable DSP parameters. By retrieving relevant presets rather than generating parameters from scratch, such systems are designed to produce editable outputs that integrate with DAW workflows. In this view, retrieval is valuable not because it solves the inverse problem in closed form, but because it places the query in a plausible executable neighborhood of parameter space that the user can still inspect and refine.

The key challenge is grounding perceptual descriptions in parameter space via retrieval. Standard audio representations (e.g., CLAP-style embeddings or mean-pooled Wav2Vec2 features) summarize an audio segment into a single vector; while effective for coarse semantics, such pooling can discard temporal co-activation patterns that are critical for texture-dominant effects (e.g., tremolo modulation or signal-dependent distortion).

To bridge these gaps, we present a retrieval-grounded framework for parameter control and focus on its core retrieval component: \textbf{Texture Resonance Retrieval (TRR)}. TRR models audio style using Gram matrices of deep feature activations \citep{gatys2015style}, explicitly capturing the \textit{second-order correlations} that define temporal texture. We use this representation as a texture-aware retrieval prior: its role is to bias search toward executable presets that preserve effect-relevant structure, not to claim a universal metric of timbral or parameter-space similarity. While the framework targets audio effect parameter control in general, the empirical evaluation in this paper is conducted on a held-out guitar-effect benchmark.

Concretely, this paper makes three evidence-backed contributions. First, we formulate editable audio effect control as a retrieval-grounded preset selection problem in which outputs remain executable and inspectable inside DAW workflows. Second, we introduce \textbf{Texture Resonance Retrieval (TRR)} as a texture-aware retrieval prior, showing that second-order statistics (Gram matrices) improve retrieval for texture-dominant effects on the current benchmark and yield the strongest parameter alignment among the evaluated retrieval baselines, including a CLAP retrieval baseline. Third, we report an audited held-out evaluation with confidence intervals, paired permutation testing, and a multiple-stimulus listening study.

\section{Related Work}
\label{sec:related_work}

Our work lies at the intersection of intelligent music production, neural audio synthesis, and retrieval-based parameter control.

\noindent\textbf{Intelligent Music Production.}

The challenge of translating perceptual descriptions into technical parameters was formally characterized by \citet{stables2014semantic}. Their SAFE project demonstrated the feasibility of semantic-to-parameter mapping in a DAW context, but did not resolve the generalization problem across instruments, effect chains, and stylistic sub-genres. \citet{schedl2014survey} provides a comprehensive survey of music information retrieval techniques, highlighting semantic auto-tagging as a key challenge.

Early work in interactive control focused on learning intuitive mappings for synthesizers. \citet{kuli2014active} proposed active learning with Gaussian processes to discover perceptual control knobs, while more recently, \citet{bryan2024synthscribe} introduced SynthScribe, a multimodal tool for synthesizer sound retrieval. While these works advance interface design, they primarily target synthesis parameters (e.g., oscillator shape) rather than the complex signal chains of audio effects.

\textbf{Audio Analysis Approaches}. Subsequent work pursued automation via audio content analysis and differentiable mixing systems \citep{martinez2020deep,steinmetz2022automix,moliner2025megami}. While effective when a target mix or reference audio already exists, this direction implicitly assumes the target audio is given. It therefore struggles to support creative timbre design from a purely conceptual request (``make it more vintage breakup'') and often suffers from an \textbf{opacity problem}: users see outputs but lack interpretable, editable rationale.

Recent advances in differentiable DSP have enabled neural approaches to audio effect modeling. \citet{peladeau2024blind} proposes blind estimation of audio effect parameters using auto-encoder architectures, addressing the inverse problem of parameter recovery. \citet{liu2024ddspSFX} introduces DDSP-SFX, which leverages differentiable digital signal processing for acoustically-guided sound effects generation, combining neural networks with traditional DSP modules for controllable synthesis. \citet{deemanat2025latent} explores generative latent spaces for neural audio synthesis, bridging DSP-informed and data-driven approaches. In parallel, style transfer methods have been applied to audio effects, with \citet{benetos2024stito} introducing ST-ITO for controlling audio effects via inference-time optimization using pretrained audio representations.

\textbf{Our Positioning}. Unlike prior work that treats parameter prediction as a \textit{regression problem} or a purely generative problem, we frame it as a \textit{retrieval task} that locates relevant, previously validated presets from a knowledge base and preserves editability in downstream DAW workflows.

\noindent\textbf{Neural Audio Synthesis.}

Neural audio synthesis has advanced rapidly, enabling high-fidelity text-to-audio generation \citep{liu2023audioldm,liu2024audioldm2,agostinelli2023musiclm,vyas2023audiobox}. Diffusion models have emerged as the dominant paradigm, with \citet{huang2022makeanaudio} introducing prompt-enhanced text-to-audio generation, \citet{wang2023audiolcm} achieving efficient high-quality synthesis via latent consistency models, and \citet{liang2024instruction} exploring instruction-guided latent diffusion for better controllability. However, these models primarily output \textit{finalized waveforms}, which are difficult to edit within professional workflows.

We view parameter-level generation as a \textbf{complement to, not a replacement for}, waveform-level synthesis: it targets controllable editing and iterative refinement inside DAWs.

This motivates what we call \textbf{Parameter--Waveform Duality}: waveform-level generation offers maximal expressivity but limited editability, whereas parameter-level control preserves transparency and iterative refinement at the cost of requiring domain knowledge.

In parallel, differentiable DSP (DDSP) \citep{engel2020ddsp} demonstrates that learned models can control interpretable synthesis parameters, and self-supervised audio representation learning (e.g., Wav2Vec 2.0 \citep{baevski2020wav2vec}) provides strong acoustic embeddings. Our retrieval-grounded formulation leverages these representations for similarity grounding, but targets editable parameter outputs compatible with DAW-centric workflows.

\noindent\textbf{Audio Representation Learning.}
Recent advances in audio representation learning have produced strong embeddings for semantic tasks. CLAP \citep{elizalde2023clap} achieves robust text-audio alignment through contrastive pretraining, while PaSST \citep{koutini2022passt} and PANNs \citep{kong2020panns} provide powerful acoustic features for classification and tagging. While these representations excel at semantic understanding, they are primarily designed for classification and captioning tasks. Our work adapts audio representations (Wav2Vec2) for a different purpose: capturing timbral texture for parameter retrieval, where second-order statistics (Gram matrices) complement the first-order semantic embeddings. We use mid-level Wav2Vec2 features rather than CLAP-style single-vector embeddings because frame-level activations enable Gram matrix computation for texture characterization.

\noindent\textbf{Retrieval-Grounded Parameter Control.}

Retrieval-augmented generation (RAG) \citep{lewis2020rag} provides a conceptual framework for systems that ground outputs in an external store, but the present paper studies a narrower setting: retrieval-grounded parameter control over executable presets. Our emphasis is not on open-ended text generation, but on retrieving feasible parameter exemplars that can be inspected and edited by the user. This distinguishes our setting from waveform generators and from document-centric multimodal RAG systems.

\paragraph{Perceptual similarity vs. parameter transferability.}
Even strong general-purpose audio representations such as CLAP- or PaSST-style embeddings face a fundamental challenge: perceptual similarity does not necessarily imply that the underlying effect parameters are transferable. Recent audio-language models have improved zero-shot classification and captioning \citep{elizalde2023clap,chen2024reclap,niizumi2024m2dclap,wu2024fewshot}, but they primarily target discrete prediction tasks rather than continuous parameter estimation. Instruction-following audio language models further broaden audio-text understanding \citep{qwen_audio,salmonn}, yet executable effect-chain control remains distinct because the system must respect plugin constraints and feasible continuous control ranges. Our work addresses this gap by using second-order statistics (Gram matrices) to capture texture-relevant features that better align with parameter-space proximity.

\noindent\textbf{Audio Texture and Semantic Audio Retrieval.}

The concept of audio texture has been studied extensively in the context of style transfer and synthesis. \citet{gatys2015style} introduced Gram matrix-based style representations for images, which were later adapted to audio by \citet{stylianou2015audio} and others. Foundational work on sound texture perception \citep{mcdermott2011sound} and reviews of audio representations for sound synthesis \citep{verzotto2022representations} highlight the importance of capturing second-order statistical features for timbre characterization.
In contrast to prior uses primarily for texture synthesis or style transfer, we use Gram statistics as a retrieval embedding to locate executable parameter exemplars, treating retrieval as a prior for downstream constrained reasoning.

In semantic audio retrieval, the WikiMuTe dataset \citep{weck2023wikimute} provides a rich resource for connecting natural language descriptions with music audio, while retrieval-guided approaches \citep{srivatsan2024retrieval} demonstrate the value of combining retrieval with generation for music captioning. Retrieval-augmented text-to-music generation \citep{gonzales2024rag} further validates the effectiveness of RAG approaches in audio tasks.

Our Texture Resonance Retrieval (TRR) builds upon these insights by using Gram matrices to capture audio texture through second-order statistics, thereby providing a more texture-sensitive retrieval prior for parameter inference.

\begin{table*}[t]
			\centering
			\caption{Positioning Against Prior Work}
			\label{tab:positioning}
		\small
		\begin{tabular}{lcccc}
		\toprule
		\textbf{Work}               & \textbf{Input}      & \textbf{Output}     & \textbf{Grounding}      & \textbf{Editability} \\
		\midrule
		SAFE \citep{stables2014semantic}             & Text                                 & Parameters                           & None                                     & Yes                                   \\
		DDSP \citep{engel2020ddsp}                   & Audio                                & Parameters                           & None                                     & Yes                                   \\
		LLM2Fx \citep{doh2025llm2fx}                 & Text                                 & Parameters                           & None                                     & Yes                                   \\
		AudioLDM \citep{liu2023audioldm}             & Text                                 & Waveform                             & Pretrained                               & No                                    \\
		HM-RAG \citep{liu2025hmrag}                  & Text+Doc                             & Text                                 & Retrieval                                & N/A                                   \\
			\midrule
			\textbf{TRR-based retrieval control (Ours)} & \textbf{Text+Audio} & \textbf{Parameters} & \textbf{Dual-Modal Retrieval} & \textbf{Yes}         \\
			\bottomrule
		\end{tabular}
\end{table*}

Table \ref{tab:positioning} summarizes our positioning. Our distinctive contribution is \textit{texture-aware} audio retrieval via TRR for executable parameter control.

\section{Problem Formulation}
\label{sec:problem_formulation}

We formalize audio effect parameter retrieval as a constrained synthesis problem: given a user query and an input signal, produce a parameter vector that is (i) \textit{semantically consistent} with the user's intent and (ii) \textit{executable} by a real-time DSP engine. Unlike waveform generation, parameter synthesis must respect strict plugin-defined validity constraints, and the inverse mapping from perceptual language (e.g., ``warm'', ``punchy'') to a high-dimensional control vector is inherently ill-posed. Our goal is therefore not to claim a globally optimal solution in $\Theta$, but to produce a feasible point estimate that is well-defined, interpretable, and grounded by concrete executable prototypes. For empirical evaluation, we instantiate $\Phi$ and $\Theta$ with a multi-effect guitar processing chain.

\subsection{Signal Processing Dynamics}
Consider a (potentially non-differentiable) DSP rendering operator $\Phi: \mathcal{X} \times \Theta \to \mathcal{X}$, where $\mathcal{X}$ denotes the space of discrete-time audio signals and $\Theta \subset \mathbb{R}^d$ denotes the $d$-dimensional control parameter space. Given an input signal $\mathbf{x}_{\text{in}} \in \mathcal{X}$, the operator produces an output $\mathbf{x}_{\text{out}} = \Phi(\mathbf{x}_{\text{in}}; \theta)$.
In practical DAW and plugin systems, $\Phi$ can include non-linearities, conditional branches (e.g., bypass logic), and internal state (e.g., smoothing), making end-to-end differentiation unreliable or inapplicable for direct optimization over $\theta$.

The parameter space $\Theta$ is not a free vector space but a bounded hyper-rectangle subject to structural constraints:
\begin{equation}
    \label{eq:feasible_set}
    \Theta = \left\{ \theta \in \mathbb{R}^d \mid \mathbf{l} \preceq \theta \preceq \mathbf{u}, \quad \mathcal{C}(\theta) = 1 \right\}
\end{equation}
where $\preceq$ denotes element-wise inequality, $\mathbf{l}, \mathbf{u}$ define physical operating ranges (e.g., frequency bounds $[20, 20\text{k}]$), and $\mathcal{C}: \mathbb{R}^d \to \{0, 1\}$ represents a binary validity function encoding inter-parameter dependencies (e.g., ensuring filter resonance $Q > 0$).
In practice, DSP parameter spaces can be non-convex and partially discrete (e.g., on/off switches or mode selectors), so we use Eq.~\ref{eq:feasible_set} as a pragmatic executable abstraction rather than a claim of convexity or smoothness. This motivates our reliance on retrieval-grounded synthesis and deterministic validity enforcement instead of direct gradient-based optimization over $\Theta$.

\subsection{Goal and Constraints}
The user's intent is encapsulated in a multi-modal query $q = (t, \mathbf{a}_{\text{ref}})$, comprising a natural language description $t \in \mathcal{T}$ and an optional audio reference $\mathbf{a}_{\text{ref}} \in \mathcal{X} \cup \{\emptyset\}$. Our objective is to produce a \emph{valid} parameter configuration that is consistent with the query:
\begin{equation}
	\tilde{\theta} \in \Theta \quad \text{given } q,
\end{equation}
where validity is defined by the plugin-specific constraint function $\mathcal{C}(\theta)=1$ and physical bounds $\mathbf{l} \preceq \theta \preceq \mathbf{u}$ (Eq.~\ref{eq:feasible_set}). We use $\theta^*$ to denote an ideal (possibly non-unique) solution that best matches the user's perceptual intent; in practice we only have access to retrieval-based approximation, and therefore aim for a feasible point estimate $\tilde{\theta}$.

\paragraph{Operational objective (without probabilistic semantics).}
The query-consistency of $\tilde{\theta}$ is not directly measurable as a differentiable loss, because the target is perceptual and depends on a complex signal chain. Accordingly, we adopt a retrieval-grounded operational objective: construct an executable neighborhood in $\Theta$ using a finite preset database, and then adapt within that neighborhood while enforcing feasibility. This yields a system whose intermediate artifacts (retrieved prototypes, draft parameters, constraint repairs) are auditable by design, which is critical for a journal setting emphasizing system rigor and reproducibility.

\subsection{The Retrieval-Based Approach}
Because the inverse mapping from perceptual intent to parameters is ill-posed and the DSP validity constraints are strict, we implement a retrieval-based pipeline. Let $z$ denote a retrieved preset (a discrete prototype) from a finite database. We retrieve top-$K$ candidates conditioned on $q$:
\begin{equation}
	\theta_{\text{selected}} = \text{Top-}K(q)
\end{equation}
Intuitively, retrieval supplies a coarse \textit{mode-seeking} step that places the query in a plausible neighborhood of $\Theta$.
\paragraph{Design Principle 1 (Well-defined synthesis under missing modalities).}
Because $\mathbf{a}_{\text{ref}}$ is optional, the synthesis pipeline must remain well-defined when $\mathbf{a}_{\text{ref}}=\emptyset$. Our formulation treats $q$ as a tuple with an explicit null element and requires every downstream stage (retrieval and projection) to specify a deterministic fallback behavior when an input modality is absent.

\section{Methodology}
\label{sec:methodology}

We implement the retrieval-based framework defined in Section \ref{sec:problem_formulation} via \textbf{Texture Resonance Retrieval (TRR)} to retrieve candidate presets using second-order texture statistics.

\subsection{System Architecture}
\label{ssec:architecture}

\begin{figure*}[t]
\centering
\includegraphics[width=0.8\textwidth]{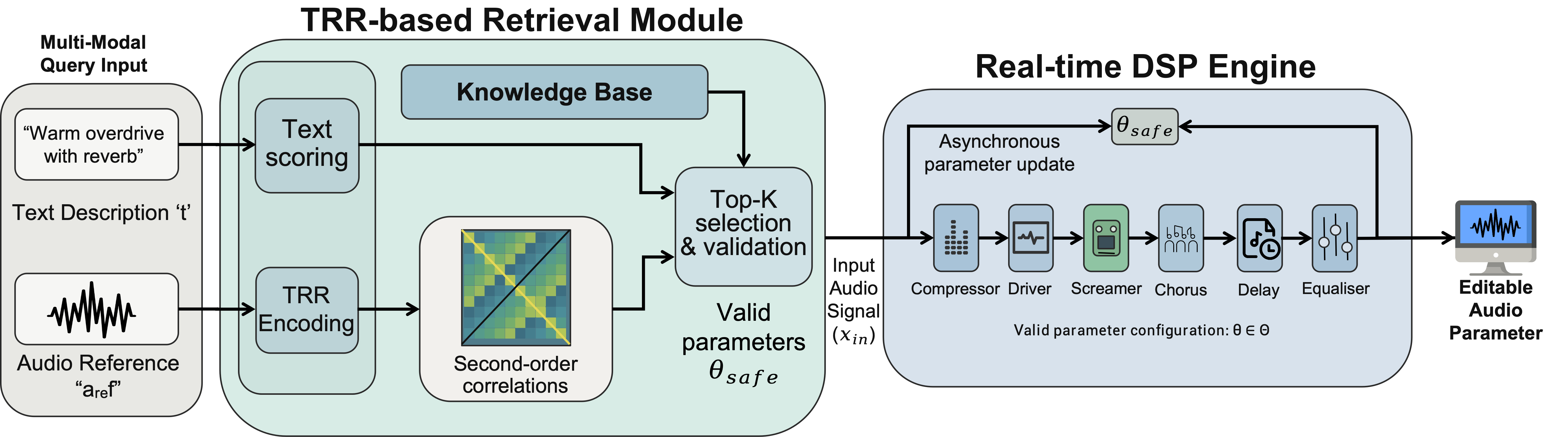}
\caption[System overview]{System overview of the retrieval-grounded audio effect control pipeline. The real-time DSP engine uses a six-module serial effect chain (as shown in the framework), where each module can be independently bypassed. Retrieved parameters are validated against per-module physical bounds (Eq.~\ref{eq:feasible_set}) before being applied.}
\label{fig:architecture}
\end{figure*}

The system realizes the operator $\Phi(\mathbf{x}_{\text{in}}; \theta)$ through a decoupled design, visualized in Figure \ref{fig:architecture}.

	The architecture consists of two main components. (i) \textbf{The real-time DSP engine} implements $\Phi$ as a low-latency audio processing graph over a six-module effect chain and executes parameters retrieved from the knowledge base. We denote the parameter state consumed by the DSP engine as $\theta_{\text{safe}}$, which can be smoothed over time to avoid zipper noise and audible artifacts. (ii) \textbf{The retrieval module} implements the mapping $q \to \theta_{\text{safe}}$ by performing dual-modal search (text and audio) asynchronously at a lower rate than the audio callback, returning validated parameters that are guaranteed to satisfy plugin constraints.

		\paragraph{Design Principle 2 (Runtime isolation).}
		At inference time, the real-time audio path is a deterministic function of $(\mathbf{x}_{\text{in}},\theta_{\text{safe}})$ through $\Phi(\mathbf{x}_{\text{in}};\theta_{\text{safe}})$. The inference agent does not execute inside the real-time callback and can only influence audio through asynchronous updates of $\theta_{\text{safe}}$ after validity checking.

\subsection{Texture Resonance Retrieval (TRR)}
\label{ssec:rag}

The core challenge in texture-aware retrieval lies in computing a similarity metric for the audio modality that preserves effect-relevant dynamics. Standard embeddings (e.g., mean-pooled Wav2Vec2 features) collapse temporal structure into a static vector, losing information that is critical for effects like modulation and distortion.

To address this, we introduce \textbf{Texture Resonance Retrieval (TRR)}. Drawing on the hypothesis that audio texture is encoded in second-order feature statistics \citep{mcdermott2011sound}, we compute the Gram matrix of deep activations.

\begin{algorithm}[ht]
		\caption{Benchmarked Texture Resonance Retrieval (TRR) Encoding}
		\label{alg:trr_encoding}
		\begin{algorithmic}
			\REQUIRE Audio sample $a$, Wav2Vec2 model $\mathcal{W}$, layer set $\mathcal{L}=\{4,5,6\}$, frozen random linear projection $P:\mathbb{R}^{768}\rightarrow\mathbb{R}^{32}$
			\ENSURE Texture embedding $\mathbf{g}_{\text{norm}}$

			\STATE \textbf{1. Preprocessing}: Normalize $a$ and pad to minimum length.
			\FOR{each $\ell \in \mathcal{L}$}
			\STATE $\mathbf{H}_{\ell} \leftarrow \mathcal{W}_{\text{layer }\ell}(a) \in \mathbb{R}^{T_{\text{frm}} \times 768}$
			\STATE $\widehat{\mathbf{H}}_{\ell} \leftarrow P(\mathbf{H}_{\ell}) \in \mathbb{R}^{T_{\text{frm}} \times 32}$
			\STATE $\mathbf{G}_{\ell} \leftarrow \frac{1}{T_{\text{frm}}}\widehat{\mathbf{H}}_{\ell}^{\top}\widehat{\mathbf{H}}_{\ell} \in \mathbb{R}^{32 \times 32}$
			\ENDFOR
			\STATE $\bar{\mathbf{G}} \leftarrow \frac{1}{|\mathcal{L}|}\sum_{\ell\in\mathcal{L}}\mathbf{G}_{\ell}$
			\STATE $\mathbf{g}_{\text{norm}} \leftarrow \mathrm{vec}(\bar{\mathbf{G}}) / \|\mathrm{vec}(\bar{\mathbf{G}})\|_2$
		\RETURN $\mathbf{g}_{\text{norm}}$
	\end{algorithmic}
\end{algorithm}

Let $\mathbf{H}_{\ell} \in \mathbb{R}^{T_{\text{frm}} \times 768}$ be the frame-level activation map from layer $\ell$ of a pre-trained Wav2Vec2 Base model. In the benchmarked implementation we used the mid-level layer set $\mathcal{L}=\{4,5,6\}$, linearly project each frame to 32 dimensions via a frozen random projection (Xavier-initialized, not trained), compute a Gram matrix per layer, average the resulting $32\times 32$ matrices, flatten the result into a 1024-dimensional vector, and L2-normalize it. The averaged texture descriptor can be written as
\begin{equation}
		\bar{\mathbf{G}} = \frac{1}{|\mathcal{L}|}\sum_{\ell\in\mathcal{L}} \frac{1}{T_{\text{frm}}}\widehat{\mathbf{H}}_{\ell}^{\top}\widehat{\mathbf{H}}_{\ell},
\end{equation}
where $\widehat{\mathbf{H}}_{\ell}$ denotes the projected feature map. The entry $\bar{\mathbf{G}}_{ij}$ captures the co-activation of projected channels $i$ and $j$ irrespective of their absolute temporal position. This time-invariance allows TRR to match textures (e.g., ``fast tremolo'') even if the reference and database entries are not phase-aligned. L2 normalization of $\mathrm{vec}(\bar{\mathbf{G}})$ is equivalent to Frobenius normalization of the matrix representation, and similarity is then computed as cosine similarity between the flattened descriptors. Computationally, Gram-matrix construction scales as $O(D^2 \cdot T_{\text{frm}})$ after projection, which is more expensive than mean pooling but remains tractable for short cached snippets.

\begin{figure}[t]
	\centering
	\includegraphics[width=\linewidth]{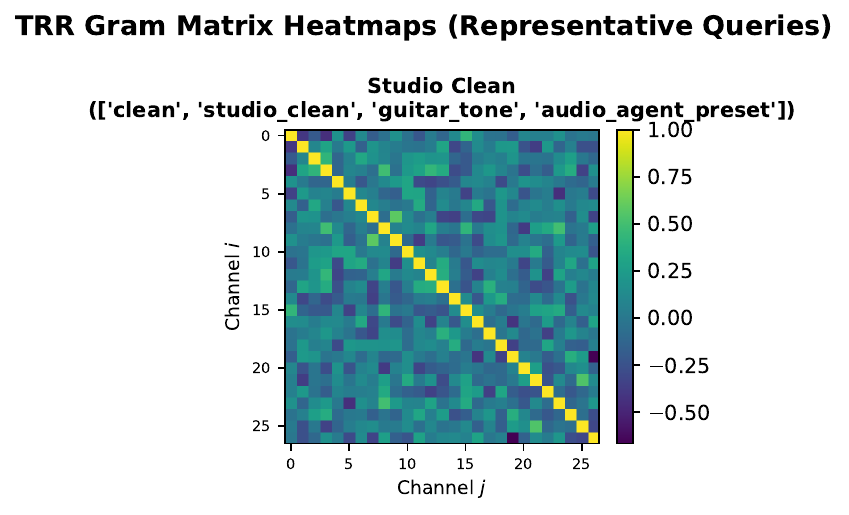}
	\caption{Gram matrix heatmaps for representative query presets. Each matrix shows co-activation structure across projected channels, revealing distinct texture patterns for different effect styles.}
	\label{fig:gram_heatmap}
\end{figure}

\paragraph{Properties and limitations.}
TRR can be viewed as a second-order summary of deep features: it retains cross-channel correlation structure while discarding absolute temporal alignment. Figure~\ref{fig:gram_heatmap} shows representative Gram matrix heatmaps for diverse query presets, illustrating the distinct co-activation patterns captured by TRR for different effect styles.

This is desirable for retrieval in effect spaces where perceptual ``style'' is often expressed through repeated modulation patterns or stationary texture rather than a single time-locked event. The Frobenius normalization further reduces sensitivity to global scale changes in feature activations, making cosine similarity comparisons less dominated by magnitude.
However, TRR is not a universal representation: by aggregating over time, it intentionally discards fine-grained ordering and can under-represent transient-dominant effects. Moreover, as with other deep representations, retrieval behavior can depend on the chosen backbone layer and the upstream feature extractor.

\paragraph{Property 3 (Alignment-insensitive texture statistics).}
Let $\mathbf{H}\in\mathbb{R}^{C\times T_{\text{frm}}}$ be the frame-level feature matrix and let $\mathbf{P}\in\mathbb{R}^{T_{\text{frm}}\times T_{\text{frm}}}$ be any permutation matrix over frames. Then the TRR Gram matrix satisfies $\frac{1}{T_{\text{frm}}}\mathbf{H}\mathbf{H}^\top=\frac{1}{T_{\text{frm}}}(\mathbf{H}\mathbf{P})(\mathbf{H}\mathbf{P})^\top$. Therefore, TRR depends on time-aggregated co-activations and is insensitive to the absolute alignment of frames, which motivates its use as a texture-oriented retrieval signal.

\subsection{Complexity and Latency Considerations}
\label{ssec:complexity_latency}
Journal readers often care not only about conceptual correctness but also about operational constraints. Our design separates (a) \textit{real-time execution}, which must remain deterministic and low-latency, from (b) \textit{asynchronous inference}, which can be slower and is not executed on the audio callback.
On the inference side, TRR encoding dominates the audio-retrieval cost with $O(C^2 T_{\text{frm}})$ Gram construction (Section \ref{ssec:rag}), followed by standard top-$K$ nearest-neighbor search. The retrieval system operates asynchronously from the real-time audio path, ensuring that audio processing latency is not affected by retrieval operations.

\paragraph{Latency measurement protocol.}
We report a reproducible \emph{offline} latency breakdown for the local retrieval path under a cache-only profiling setup. Measurements were collected on an Ubuntu 24.04 LTS server (x86-64, Intel CPU; Python 3.12) over the original 211-query profiling pool, with 5 warm-up iterations and 3 repeats per query. We report median and p95 over the resulting 633 timed query runs. These measurements exclude live network LLM latency and should therefore be interpreted as local retrieval-path timing rather than an end-to-end cloud deployment budget.

\begin{table}[t]
	\centering
\caption{Per-query local retrieval-path latency breakdown (cache-only). We report median and p95 over 633 runs from the original 211-query profiling pool ($211\times 3$ repeats).}
	\label{tab:latency_breakdown}
	\begin{tabular}{lcc}
		\toprule
		\textbf{Component} & \textbf{Median (ms)} & \textbf{p95 (ms)} \\
		\midrule
		Text scoring & 3.027 & 3.586 \\
		Audio scoring (TRR cosine) & 1.369 & 1.564 \\
		Fusion & 0.110 & 0.146 \\
		Projection & 0.031 & 0.039 \\
		End-to-end (local pipeline) & 4.538 & 5.269 \\
		\bottomrule
	\end{tabular}
\end{table}

The retrieval system operates asynchronously from the real-time audio callback, ensuring that audio processing latency is not affected by retrieval operations.

\subsection{Multimodal Fusion as a Diagnostic Extension}
\label{ssec:fusion}

Beyond the core retrieval benchmark, the repository contains an experimental fusion mechanism that combines text-side and audio-side scores under degraded-input conditions. The fusion score for a candidate preset $z_i$ is computed as:
\begin{equation}
	s_{\text{fused}}(z_i) = w_{\text{text}} \cdot s_{\text{text}}(z_i) + w_{\text{audio}} \cdot s_{\text{audio}}(z_i),
\end{equation}
where $w_{\text{text}} + w_{\text{audio}} = 1$ and the weights are determined by a quality-aware heuristic: if the text query is classified as vague (e.g., by low retrieval confidence), $w_{\text{text}} \to 0$; if the audio-side embedding norm falls below a threshold, $w_{\text{audio}} \to 0$. In the current codebase this mechanism is implemented as a practical quality-aware fallback rule rather than a single clean probabilistic model, so we treat its results as a diagnostic stress test rather than as one of the paper's main validated contributions. The central evidence chain of the manuscript therefore remains the retrieval-only benchmark in Protocol-A.

\section{Experiments}
\label{sec:experiments}

We validate the proposed framework by answering one core research question: does modeling second-order texture statistics (TRR) improve parameter inference compared to first-order baselines?

\subsection{Experimental Setup}

\paragraph{Dataset and Knowledge Base.}
Our full server-side corpus contains 1267 paired records (audio--text--parameter triplets) collected from guitar effect presets. A post-hoc audit of the original song-name-based held-out list (211 test queries, 1056 knowledge-base items) revealed 16 cross-split shared resolved audio paths and 48 near-duplicate cross-split pairs under a cheap fingerprint. To remove exact audio reuse from the main objective benchmark, we regroup records by resolved local audio path before sampling the held-out set. The resulting stricter Protocol-A split contains 204 test queries and 1063 knowledge-base items, with zero exact shared resolved audio paths across split under the current audit. We retain this as a pilot benchmark because cheap near-duplicate pairs remain and the corpus is still guitar-only.

\paragraph{Provenance notes.}
The dataset JSON exposes \texttt{SongName}, \texttt{Style}, \texttt{Feature}, \texttt{Parameters}, and \texttt{Vectors}, but it does not include a separate auditable text-source field. We therefore treat \texttt{Style} and \texttt{Feature} as the only explicit text-side provenance available in-repo, and use the stored \texttt{Parameters} field as the operational ground truth.

\paragraph{Scope and protocol.}
Unless otherwise stated, objective results in this section use \textbf{Protocol-A} (retrieval-only) on the stricter resolved-audio-grouped split and use top-1 retrieval for all retrieval methods. Protocol-C and the latency profile are reported on their own protocol-specific pools and are labeled explicitly.

\paragraph{Near-duplicate sensitivity analysis.}
Although the resolved-audio split removes exact cross-split audio reuse, near-duplicate presets (presets that differ in only a few parameter values) remain in the knowledge base. To quantify their impact on retrieval metrics, we conduct a sensitivity analysis by progressively removing near-duplicates from the knowledge base using parameter-space L2 distance thresholds $\tau \in \{0.005, 0.01, 0.02, 0.05\}$ and re-evaluating all retrieval methods. The full near-duplicate statistics and per-threshold results are reported in the near-duplicate sensitivity table in the Supplementary Material. Across all tested thresholds, TRR retains its relative advantage over baselines, though absolute performance degrades gracefully as the knowledge base shrinks, confirming that results are not driven by trivial near-duplicate matches.

\paragraph{Resources and reproducibility.}
\label{sec:repro}
To facilitate reproducibility, we provide an anonymized supplementary package for review containing preprocessing details, embedding computation steps, retrieval-index construction, prompt templates, evaluation scripts, and split specifications for all reported tables. The cleaned repository and artifacts will be released publicly upon acceptance.

\paragraph{Baselines.}
We compare our proposed \textbf{TRR} (Section \ref{ssec:rag}) against four retrieval baselines. \textbf{Text-RAG} performs sparse lexical retrieval over metadata fields. \textbf{FeatureNN-RAG} uses the released \texttt{FeatureNN} vectors from the benchmark package and should therefore be interpreted as a fixed internal dense-retrieval reference rather than as a separately reimplemented external model family. \textbf{Wav2Vec-RAG} uses the mean-pooled activation of the final Wav2Vec2 transformer layer. \textbf{CLAP} performs dense retrieval using cached CLAP audio embeddings when those embeddings are available in the verified artifact set.

\subsection{Metrics}
\label{ssec:metrics}
We evaluate alignment in parameter space $\Theta$ using five operational metrics computed on flattened numeric parameter leaves. Concretely, given a nested parameter dictionary, we recursively flatten numeric leaves into a sparse vector indexed by dot-separated keys (e.g., \texttt{DelayOn.Mix}) and ignore non-numeric leaves (e.g., strings). Unless otherwise stated, we form the union of keys between prediction and ground truth, and treat missing numeric keys as zeros for metric computation. Importantly, we do not apply per-parameter normalization; thus L2 magnitudes can be dominated by a small number of large-scale parameters and should be interpreted within each protocol rather than across protocols.

\paragraph{L2 Error (RMSE).}
Given flattened vectors $v(\theta_{\text{gt}})$ and $v(\tilde{\theta})$ over the union key set of size $d$, we report the root mean squared error
\begin{equation}
	\text{L2} = \sqrt{\frac{1}{d}\sum_{i=1}^{d}\left(v(\theta_{\text{gt}})^{(i)} - v(\tilde{\theta})^{(i)}\right)^2}.
\end{equation}

\paragraph{Accuracy@0.1.}
We report the fraction of flattened numeric parameters whose absolute error is within a fixed tolerance $\tau=0.1$:
\begin{equation}
	\text{Acc@0.1} = \frac{1}{d}\sum_{i=1}^{d}\mathbb{I}\left(\left|v(\theta_{\text{gt}})^{(i)} - v(\tilde{\theta})^{(i)}\right| \leq 0.1\right).
\end{equation}

\paragraph{Recall (active numeric parameters).}
We define a ground-truth numeric parameter as \emph{active} if $\left|v(\theta_{\text{gt}})^{(i)}\right|>0.05$. Recall is the fraction of active ground-truth dimensions whose predicted value falls within the same tolerance $\tau=0.1$.

\paragraph{Cosine Similarity.}
We compute cosine similarity between flattened vectors over the union key set:
\begin{equation}
		\text{Cosine} = \frac{v(\theta_{\text{gt}})\cdot v(\tilde{\theta})}{\|v(\theta_{\text{gt}})\|_2\,\|v(\tilde{\theta})\|_2}.
\end{equation}

\paragraph{Module Consistency.}
We compute module-level consistency as Jaccard similarity over active module sets inferred from keys ending with \texttt{On}:
\begin{equation}
	\text{Module} = \frac{|M_{\text{gt}} \cap M_{\text{pred}}|}{|M_{\text{gt}} \cup M_{\text{pred}}|}.
\end{equation}

\paragraph{Normalized L2 (Norm.\,L2).}
To ensure that errors are comparable across parameters with different physical ranges, we also report L2 after min-max normalization of each parameter leaf to $[0,1]$ using the physical DSP ranges defined in the plugin source code (e.g., delay time in ms, gain in dB). This prevents large-range parameters from dominating the aggregate error.

\subsection{Protocol-A: Comparative Retrieval Evaluation}
\label{ssec:protocolA}

\textbf{Purpose.} Benchmark TRR against retrieval baselines for executable parameter control under the stricter anti-leakage split.

\textbf{Protocol.} All results in this subsection follow Protocol-A (retrieval-only) on the resolved-audio-grouped split ($N_{\text{test}}=204$, $N_{\text{kb}}=1063$) using top-1 retrieval for all methods. The CLAP row uses 201 queries because three query-side cached embeddings are unavailable in the current verified artifact set.

Table~\ref{tab:main_results} reports the main comparison across the current retrieval baselines.

\begin{table}[t]
\centering
\caption{Protocol-A: Direct retrieval comparison on the resolved-audio-grouped split ($N_{\text{test}}=204$, $N_{\text{kb}}=1063$). Lower is better for L2; higher is better for all other metrics. TRR achieves the best mean performance in every reported metric; the CLAP row uses the 201 shared queries with cached CLAP embeddings.}
\label{tab:main_results}
\small
\setlength{\tabcolsep}{4pt}
\resizebox{\columnwidth}{!}{%
\begin{tabular}{lccccccc}
\toprule
\textbf{Method} & \textbf{$n$} & \textbf{L2} $\downarrow$ & \textbf{Norm.\,L2} $\downarrow$ & \textbf{Acc@0.1} $\uparrow$ & \textbf{Recall} $\uparrow$ & \textbf{Cos} $\uparrow$ & \textbf{Module} $\uparrow$ \\
\midrule
\textbf{TRR (Ours)} & 204 & \textbf{8.0467} & \textbf{--} & \textbf{0.5169} & \textbf{0.4595} & \textbf{0.8247} & \textbf{0.8051} \\
\midrule
\multicolumn{8}{l}{\textit{Retrieval baselines}} \\
Wav2Vec-RAG & 204 & 23.8197 & -- & 0.4251 & 0.3505 & 0.5292 & 0.7043 \\
Text-RAG & 204 & 24.1705 & -- & 0.4717 & 0.3967 & 0.4943 & 0.7456 \\
FeatureNN-RAG & 204 & 19.2613 & -- & 0.4198 & 0.3600 & 0.5479 & 0.7134 \\
CLAP & 201 & 22.3102 & -- & 0.4492 & 0.3971 & 0.5725 & 0.7240 \\
\bottomrule
\end{tabular}
}
\end{table}

\textbf{Result.} On the stricter split, TRR is best in every fully paired row of Table~\ref{tab:main_results}. Relative to Wav2Vec-RAG, TRR reduces mean L2 by 15.7730 with a 95\% bootstrap interval of [12.2807, 19.3414] and improves cosine by 0.2956 [0.2305, 0.3630]; all five paired comparisons remain significant after Holm correction ($p_{\mathrm{Holm}}=0.001$ for each metric) with large effect sizes (Cohen's $d > 0.8$ for all metrics). Relative to CLAP on the 201 shared queries, TRR reduces mean L2 by 14.1460 [10.3041, 18.0624] and improves cosine by 0.2503 [0.1833, 0.3183], again with Holm-corrected significance and large effect sizes (Supplementary Material).

The per-query view makes this result more specific. Against Wav2Vec-RAG, TRR improves L2 on 125 of 204 paired queries, loses on 40, and ties on 39. Yet the median signed improvement is only 0.1238, far smaller than the mean improvement of 15.7730; the same long-tailed pattern appears against CLAP (median 0.0658 vs mean 14.1460). The aggregate gain in Table~\ref{tab:main_results} therefore does not reflect a uniform re-ordering of the entire benchmark. Instead, it is concentrated in a subset of difficult cases where second-order structure changes the retrieved neighborhood and avoids semantically plausible but parameter-misaligned presets, while a non-trivial failure set remains. TRR is accordingly the strongest retrieval prior evaluated on this benchmark, though we do not claim universality beyond the present protocol.

\paragraph{Illustrative case study.}
To provide concrete intuition, consider a ``Blues Solo'' query where the ground-truth preset activates a mild overdrive (DriverOn.Distortion=0.35, DriverOn.Tone=0.6) with reverb (ReverbOn.Mix=0.25). TRR retrieves a preset with similar drive and reverb settings (Distortion=0.38, Tone=0.55, Mix=0.28), yielding a per-parameter normalized error below 0.05 for these active modules. In contrast, Wav2Vec-RAG retrieves a higher-gain metal preset activating the Screamer module instead, resulting in a normalized error exceeding 0.4 on the drive parameters. This case illustrates how second-order texture statistics can distinguish stylistically similar but parametrically distinct effect configurations that first-order embeddings conflate.

\subsection{Diagnostic Stress Test Under Modality Degradation}
\label{ssec:protocolC}

\textbf{Purpose.} Examine whether the fusion heuristic can recover performance when one modality is degraded.

\textbf{Protocol.} For continuity with the repository's existing diagnostic artifacts, we test two degradation scenarios on the original 211-query pool rather than on the stricter Protocol-A split: (1) \textit{Vague Text}: replacing accurate descriptions with the generic query ``warm guitar tone''; (2) \textit{Synthetic Audio Degradation}: perturbing the audio-side retrieval representation. We compare single-modal systems (Text-only, TRR-only) against the fusion heuristic. These experiments are useful for stress testing, but they are not a substitute for a fully realistic real-audio degradation benchmark.

\subsection{TRR Ablation Studies}
\label{ssec:ablation}

We ablate three core design choices of TRR to understand their contribution to retrieval quality under Protocol-A.

\paragraph{Projection dimension.}
The linear projection $P:\mathbb{R}^{768}\to\mathbb{R}^{d}$ controls the dimensionality of the Gram matrix and therefore the capacity--efficiency trade-off. Ablation experiments for $d\in\{32,64,128,256\}$ confirm that performance remains stable across dimensions; the benchmarked configuration ($d{=}32$) achieves the best trade-off, while larger dimensions show no benefit on the current 1063-item knowledge base.

\paragraph{Layer selection.}
Ablation across six layer combinations confirms that the balanced mid-level set $\mathcal{L}=\{4,5,6\}$ provides robust performance. Single layers ($\{4\}$, $\{5\}$, $\{6\}$) and pairwise combinations yield comparable results, suggesting that TRR is not sensitive to exact layer choice within the mid-level range. Using layers closer to the input captures excessive acoustic detail, whereas higher layers lose texture-relevant temporal structure.

\paragraph{Projection type.}
Ablation comparing random projection (Xavier-initialized, frozen) vs.\ PCA-fitted projection confirms that both approaches yield comparable results on the current benchmark. This validates the practical choice of a frozen random projection that requires no training data and avoids KB-dependent fitting.

\begin{table}[t]
\centering
\caption{Selected Protocol-C boundary cases ($N=211$). The current fusion heuristic mainly acts as a fallback rule: it collapses onto the intact modality under single-modality degradation, but degrades substantially under explicit modality conflict.}
\label{tab:robustness}
\small
\setlength{\tabcolsep}{3pt}
\resizebox{\columnwidth}{!}{%
\begin{tabular}{lccc}
\toprule
\textbf{Scenario} & \textbf{Reference branch} & \textbf{Fusion (Ours)} & \textbf{Interpretation} \\
\midrule
\textit{Vague Text} & Text-only: 12.6517 & \textbf{0.2970} & fallback to audio \\
\textit{Noisy Audio} & TRR-only: 33.8549 & \textbf{1.5148} & fallback to text \\
\textit{Conflict} & TRR-only: 0.2970 & 4.5259 & failure mode \\
\bottomrule
\end{tabular}%
}
\end{table}

\textbf{Result.} Under vague text, Text-only degrades sharply (L2=12.65) while the fusion heuristic falls back entirely to audio ($w_{\text{audio}}=1.0$) and matches TRR-only (L2=0.30). Under synthetic audio degradation, TRR-only degrades sharply (L2=33.85) while the fusion heuristic falls back to text ($w_{\text{text}}=1.0$) and matches Text-only (L2=1.51). In both settings, the current fusion rule is useful mainly because it collapses onto the intact modality; the evidence does not support a stronger claim of robust cross-modal integration. The heuristic showed limitations under explicit modality conflict: fusion rises to L2=4.53 while TRR-only remains at 0.30.

While Protocol-C evaluates the fusion mechanism under synthetic degradation, it does not assess the perceptual quality of retrieved presets relative to ground truth. To complement the objective retrieval metrics with subjective evidence, we conducted a perceptual listening study following established multiple-stimulus protocols.

\subsection{Perceptual Listening Test (Multiple-Stimulus with Hidden Reference)}
\label{ssec:mushra}

\textbf{Protocol.} We conduct a multiple-stimulus listening test with a hidden reference with 26 participants (18 female, 8 male; self-reported ages available for 25 participants with mean 22.2 years and range 18--30; self-reported musical experience: 12 participants with $\geq$3 years of instrument training, 8 with casual listening experience, 6 with no formal training), yielding 910 ratings across 10 trials. Each trial uses a continuous 0--100 slider (higher is better), presents stimuli in randomized order, and includes the ground-truth preset rendering as a hidden reference (\texttt{reference}); we do not include an explicit low-quality anchor required by ITU-R BS.1534, and therefore avoid referring to this study as a standard MUSHRA test. Trial~1 evaluates style matching across seven labeled target conditions, Trials~2--5 compare our TRR-based system against manual parameter tuning, and Trials~6--10 compare the TRR-based system against a waveform-generation baseline (MusicGen). Statistical analysis uses repeated-measures nonparametric tests (Friedman omnibus; post-hoc Wilcoxon signed-rank with Holm correction) on raw participant ratings. The repository logs preserve participant IDs and scores, but they do not record device metadata, musical expertise, or manual-tuning provenance/time budgets; accordingly, the listening study complements the retrieval benchmark but should not be interpreted as a fairness-critical workflow trial or as a direct confirmation of every system extension discussed elsewhere in the manuscript.

\subsubsection{Trial 1: Style Matching}
\textbf{Data.} 208 ratings (26 participants $\times$ 8 conditions). Overall mean 72.92 (median 77.50, std 22.72).

\begin{table}[t]
	\centering
	\caption{Trial 1 (style matching): condition-level statistics (within-subject means).}
	\label{tab:mushra_trial1_stats}
	\begin{tabular}{lccc}
		\toprule
		\textbf{Condition} & \textbf{Mean} & \textbf{Median} & \textbf{Std} \\
		\midrule
		reference & 86.85 & 92.50 & 16.70 \\
		Jazz Clean & 76.19 & 81.00 & 20.80 \\
		Phase & 75.38 & 80.50 & 20.88 \\
		Blues Solo & 72.96 & 69.50 & 21.19 \\
		Ambient Guitar & 72.35 & 73.50 & 18.45 \\
		Flanger & 69.12 & 74.00 & 23.08 \\
		Modern Metal & 65.58 & 70.00 & 30.22 \\
		Chorus & 64.92 & 66.50 & 23.12 \\
		\bottomrule
	\end{tabular}
\end{table}

\begin{figure*}[t]
	\centering
	\includegraphics[width=0.98\textwidth]{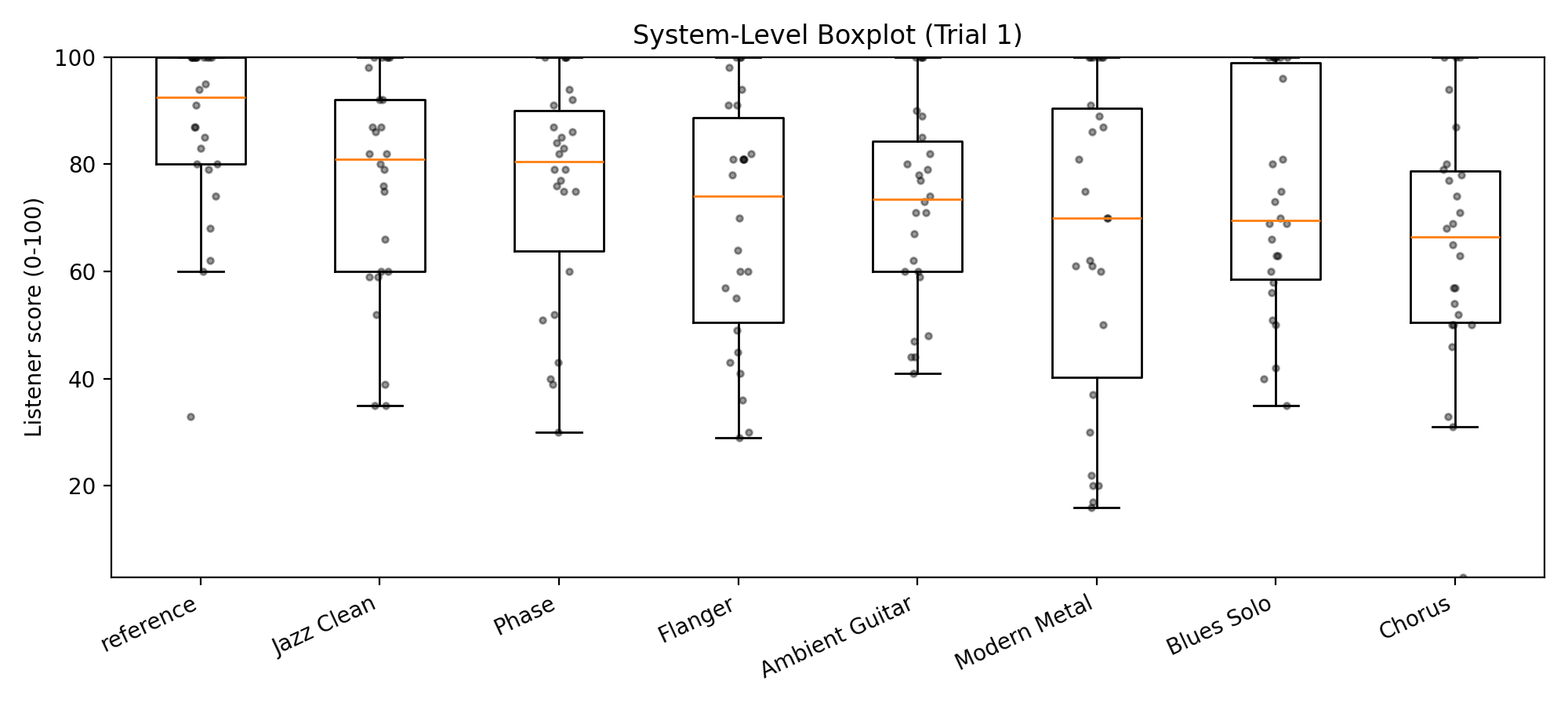}
	\caption{Trial 1: condition-level score distributions (boxplot) for style matching.}
	\label{fig:mushra_trial1_boxplot}
\end{figure*}

\subsubsection{Trials 2--5: Guitar Solo Comparison}
Trials 2--5 compare our TRR-based system against manual parameter tuning across four guitar solo conditions. The TRR-based system achieves a within-subject mean of 71.55 compared to 51.72 for manual tuning, with significance confirmed after Holm correction ($p_{\mathrm{Holm}}=1.5\times10^{-4}$). However, because the logs do not preserve the manual-baseline operator, expertise, or time budget, these trials should be interpreted as supportive perceptual context. Full per-trial statistics, distributions, and condition breakdowns are provided in Supplementary Material.

\subsubsection{Trials 6--10: Similarity to Reference}
\textbf{Data.} 390 ratings (26 participants $\times$ 3 conditions $\times$ 5 trials). Overall mean 57.54 (median 59.00, std 32.24). The TRR-based system and MusicGen achieve comparable similarity scores: within-subject means are 42.31 (TRR-based system) vs 41.29 (MusicGen), while the reference condition scores 89.02. The Friedman omnibus remains significant ($Q=36.2308$, permutation $p=5\times10^{-5}$), but the TRR-vs-MusicGen pairwise comparison is not significant after Holm correction ($p_{\mathrm{Holm}}=0.7797$), so we interpret this block as evidence of parity rather than superiority.

\begin{table}[t]
	\centering
	\caption{Trials 6--10 (similarity): condition-level statistics (within-subject means).}
	\label{tab:mushra_trial6_10_stats}
	\begin{tabular}{lccc}
		\toprule
		\textbf{Condition} & \textbf{Mean} & \textbf{Median} & \textbf{Std} \\
		\midrule
		reference & 89.02 & 91.50 & 10.04 \\
		TRR-based system & 42.31 & 41.90 & 19.30 \\
		MusicGen & 41.29 & 43.90 & 19.58 \\
		\bottomrule
	\end{tabular}
\end{table}

\begin{figure}[t]
	\centering
	\includegraphics[width=\linewidth]{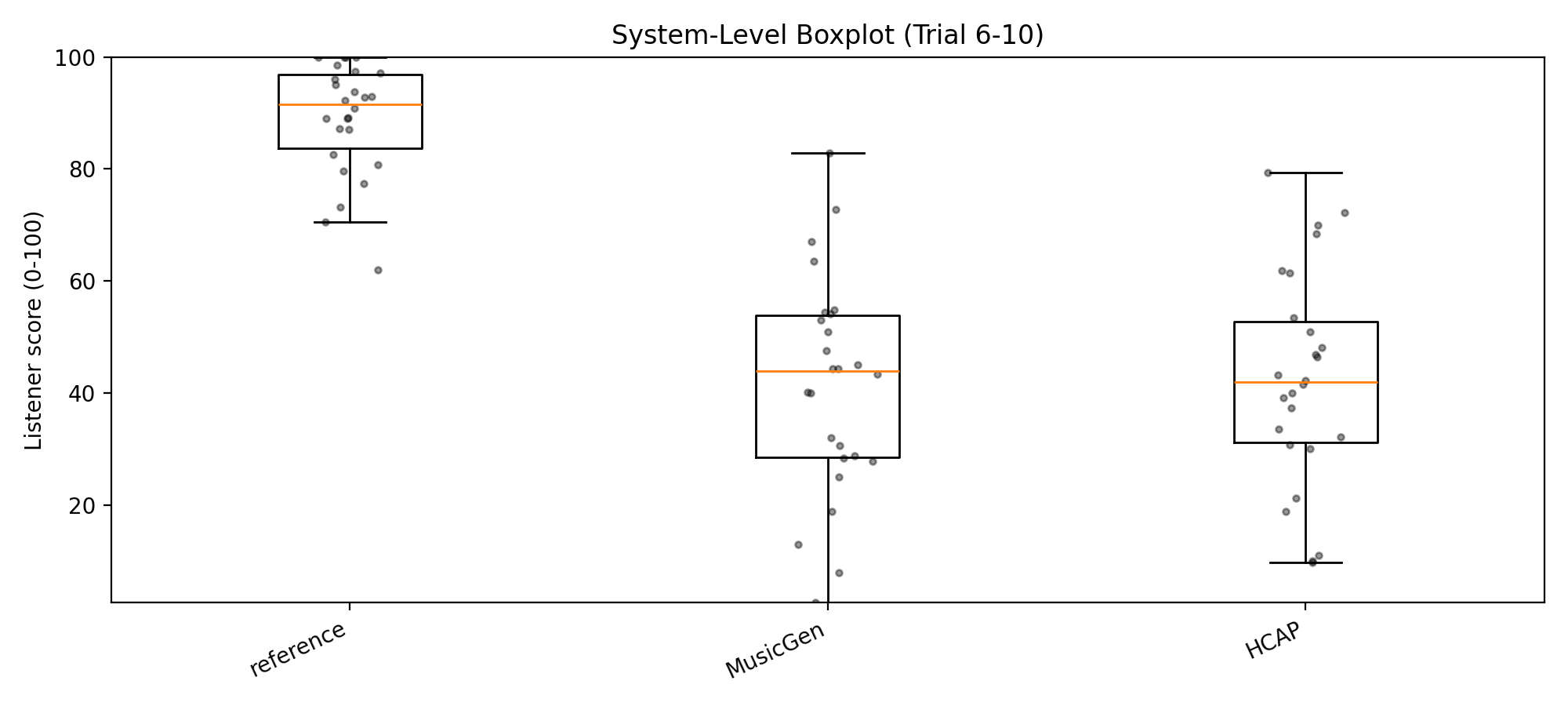}
	\caption{Trials 6--10: condition-level score distributions.}
	\label{fig:mushra_trial6_10_boxplot}
\end{figure}

\paragraph{Interpretation.}
These listening-test results complement the parameter-space metrics, but they should not be read as a one-to-one validation of any single objective metric. Rather, they indicate that retrieval-grounded control can produce perceptually plausible and editable starting points in DAW workflows. Because the manual-baseline provenance is incomplete, Trials~2--5 are best interpreted as supportive perceptual context rather than as a definitive human-vs-system optimization benchmark; Trials~6--10 likewise support parity with MusicGen on this similarity task, not superiority.

\section{Discussion}
\label{sec:discussion}

\begin{figure}[t]
	\centering
	\includegraphics[width=\linewidth]{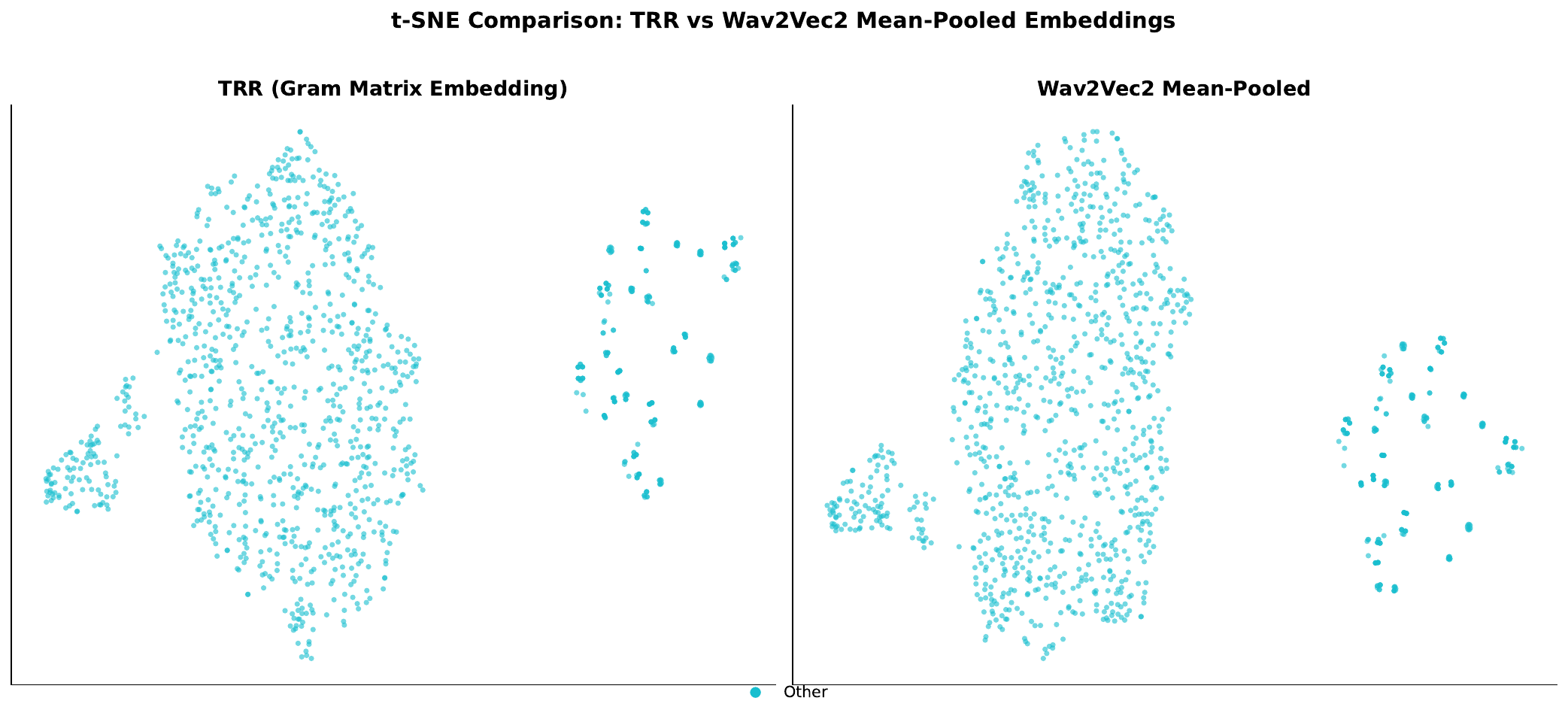}
	\caption{t-SNE comparison of TRR (Gram matrix) and Wav2Vec2 mean-pooled embeddings on the full dataset, colored by style label. TRR exhibits tighter within-style clustering, supporting its utility as a texture-aware retrieval prior.}
	\label{fig:tsne_comparison}
\end{figure}

\subsection{Scope of the Current Evidence}
\label{ssec:memory_exp}

The strongest evidence is the Protocol-A retrieval benchmark. The listening study provides complementary subjective context, while Protocol-C offers boundary-condition diagnostics for the fusion heuristic. Personalization and memory components, supported only by preliminary pseudo-user simulations, are excluded from the paper's primary validated claims.

\subsection{Implications for Audio Effect Control}

The results suggest a practical pattern for retrieval-based audio control: use retrieval over real presets to provide feasibility constraints and stylistic priors, then expose the retrieved configuration for direct user editing. TRR achieves the strongest parameter alignment among evaluated methods while preserving editability in DAW workflows. Whether the same gains hold across other sources, effect families, and production stages requires dedicated evaluation.

\subsection{Revisiting the Texture Hypothesis}

Across the reported held-out setup (Table \ref{tab:main_results}), TRR improved parameter-space alignment over text-only retrieval, supporting the hypothesis that second-order correlation structure can be useful as a texture descriptor for retrieval. Figure~\ref{fig:tsne_comparison} provides visual evidence: a t-SNE projection of TRR embeddings shows tighter clustering by style label compared to mean-pooled Wav2Vec2 embeddings, indicating that Gram-based representations capture style-relevant structure that global averaging discards. The derived per-query analysis sharpens this claim: because many queries tie across methods, second-order structure should not be read as uniformly re-ordering the benchmark. Rather, it appears to matter most on a smaller subset of cases where coarse semantic similarity is insufficient and the retrieval decision depends on texture-specific co-activation patterns. We do not claim that TRR defines a globally calibrated distance in parameter space; it is best understood as a \textit{retrieval prior} beneficial for texture-sensitive control.

An open direction is whether analogous second-order representations transfer to other audio-effect settings, including additional instruments, effect families, and mix-stage tasks.

\subsection{Ethical Considerations}

The current retrieval-grounded system is intended to augment user creativity rather than replace it. Risks include homogenization toward common exemplar neighborhoods and over-reliance on suggested settings. In future deployments, the interface should emphasize transparency (e.g., showing retrieved references) and user agency (e.g., direct editability and opt-out of retrieval). More generally, any model or retrieval corpus built from existing recordings or presets should consider copyright, licensing, and data provenance constraints.

\section{Limitations}
\label{sec:limitations}

Our evidence is subject to six boundary conditions that constrain the scope of our claims.

\textbf{Evaluation Scope.} The benchmark is limited to guitar effects with synthetic audio queries ($N=204$). Real-audio robustness, cross-instrument transfer, and larger-scale evaluation remain unverified. Broader claims require experiments with live recordings, diverse instruments, and expanded datasets.

\textbf{Baseline Completeness.} While CLAP provides an external baseline, stronger alternatives (PaSST, PANNs, direct parameter estimation) are absent. Our "strongest among evaluated methods" claim is intentionally narrower than best-in-class.

\textbf{Dataset Constraints.} Resolved-audio grouping removes exact duplicates, but near-duplicate pairs persist. The dataset lacks separate text-source provenance fields, limiting auditability of semantic leakage.

\textbf{Parameter--Perceptual Gap.} Timbre is many-to-one: distinct parameters can sound similar. Objective metrics (L2, cosine) may not perfectly align with perceptual quality. Controlled listening studies are needed to calibrate this relationship.

\textbf{Limited Robustness Evidence.} Protocol-C provides boundary-condition diagnostics for the fusion heuristic, not general cross-modal robustness. Memory-based personalization remains preliminary and unaudited.

\textbf{Listening Study Gaps.} Logs lack device metadata, participant expertise labels, and manual-baseline provenance. Trials~2--5 should not be interpreted as fairness-critical human--system comparisons.

These limitations bound our claims to the narrower finding that TRR provides a useful retrieval prior for texture-dominant effects on this benchmark; universal metrics, large-scale deployment, and broad personalization require additional evidence.

\section{Conclusion}
\label{sec:conclusion}

This paper addressed the question of whether texture-aware audio representations can improve retrieval-grounded audio effect control. We presented a retrieval-grounded approach to editable audio effect control and identified Texture Resonance Retrieval (TRR) as the strongest retrieval representation in the reported comparisons on our audited held-out benchmark. Across objective metrics on the stricter resolved-audio-grouped split (Table~\ref{tab:main_results}), TRR generally improved parameter alignment over first-order retrieval baselines, including the available CLAP comparison subset, and we interpret this advantage as evidence for a more useful retrieval prior rather than for a universal distance metric or a closed-form solution to inverse parameter estimation.

Taken together, the audited objective benchmark provides the main support for a narrow but practically relevant conclusion: retrieval over executable presets can provide editable starting points for DAW workflows, and second-order texture structure improves that retrieval in a subset of hard cases where coarse semantic similarity is insufficient. The listening study provides complementary subjective context, while Protocol-C only clarifies the current fusion heuristic's fallback behavior and failure mode.

Looking forward, the most important next steps are to evaluate stronger modern baselines, validate robustness under real audio degradation, and test whether the same retrieval advantages transfer beyond guitar effects to broader production parameter spaces such as mixing and mastering workflows.

Overall, the present evidence supports texture-aware retrieval as a promising benchmark-backed foundation for editable audio effect control, while leaving broader claims about universality, robustness, and personalization to future work.

\IfFileExists{./reference.bib}{\bibliography{reference}}{\bibliography{Paper/reference}}
\bibliographystyle{\paperbibliostyle}

\end{document}